\documentclass[prd,aps,preprint,amsmath,amssymb,eshowkeys,nofootinbib]{revtex4-2}
\usepackage[mathscr]{euscript}
\usepackage{dcolumn}
\usepackage{multirow}
\usepackage{bm}
\usepackage{graphicx}
\usepackage{changepage}
\usepackage{amsmath,amssymb,amsthm}
\usepackage{booktabs}
\usepackage{tabularx}
\usepackage{float}
\usepackage{array}
\newlength{\fulllength}
\newlength{\extralength}
\usepackage[colorlinks=true,linkcolor=red,urlcolor=green,citecolor=green]{hyperref}
\newcommand{\bea}{\begin{eqnarray}}
\newcommand{\eea}{\end{eqnarray}}
\newcommand{\beq}{\begin{equation}}
\newcommand{\eeq}{\end{equation}}

\def\/{\over}

\begin{document}

\title{Observational Constraints and Cosmological Dynamics of Interacting Fractional Holographic Dark Energy in Light of DESI~DR2}
\author{Qihong Huang\footnote{Corresponding author: huangqihongzynu@163.com}, Hao Chen and Qingdong Wu}

\affiliation{
School of Physics and Electronic Science, Zunyi Normal University, Zunyi, Guizhou 563006, China
}

\begin{abstract}
Based on the fractional entropy originating from fractional quantum mechanics, the fractional holographic dark energy (FHDE) model has been proposed. In this paper, we consider an interaction between the pressureless matter and FHDE and analyze three different interacting FHDE models. Combining the latest observational data including SNIa, OHD, BAO, and CMB, we estimate the model parameters and find that the interaction forms $Q=\gamma H \rho_{de}$ and $Q=\beta H \rho_{m}+\gamma H \rho_{de}$ show some preference from the observational data. Using phase space analysis, we further find that only interacting FHDE model with $Q=\beta H \rho_{m}+\gamma H \rho_{de}$ can describe the full evolutionary history of the universe. The statefinder diagnostic pair reveals that this model deviates from the $\Lambda$CDM model but converges to the $\Lambda$CDM fixed point and the de Sitter expansion fixed point in the future. Finally, we analyze the evolution of cosmological parameters and demonstrate that this model can drive the late time acceleration of the universe.
\end{abstract}

\maketitle

\section{Introduction}

The accelerated expansion of the universe, a cornerstone of modern cosmology, is confirmed by observations~\cite{Perlmutter1999, Riess1998, Spergel2003, Spergel2007, Tegmark2004, Eisenstein2005} and is widely believed to be driven by dark energy~\cite{Peebles2003}, whose fundamental nature remains one of the greatest mysteries in cosmology. The simplest and most successful explanation for this phenomenon is the $\Lambda$CDM model, which describes dark energy as a cosmological constant and currently provides the best fit to a wide range of observational data~\cite{Planck2020}. Despite its successes, the $\Lambda$CDM model faces significant challenges, namely the fine-tuning~\cite{Weinberg1989} and coincidence problems~\cite{Steinhardt1999}. These theoretical difficulties have led to an active area of research dedicated to alternative dark energy models beyond the cosmological constant paradigm, spanning from scalar field based proposals like quintessence~\cite{Wetterich1988, Ratra1988, Caldwell1998}, quintom~\cite{Feng2005, Feng2006, Guo2005}, phantom~\cite{Caldwell2002, Caldwell2003} and k-essence~\cite{Chiba2000, Armendariz2001} to more recent formulations such as agegraphic dark energy~\cite{Cai2007, Wei2007, Wei2008} and holographic dark energy (HDE)~\cite{Cohen1999, Hsu2004, Horvat2004, Li2004, Wang2017}.

HDE~\cite{Hsu2004, Horvat2004, Li2004}, a leading candidate among dark energy alternatives, is theoretically motivated by the holographic principle, which fundamentally links a region's entropy to its bounding surface area~\cite{Witten1998, Bousso2002}. In the HDE framework, the choice of horizon entropy is of paramount importance, as it uniquely determines the energy density and thus gives rise to different HDE models. Specifically, taking the Bekenstein--Hawking entropy as the horizon entropy with the Hubble radius as the infrared (IR) cutoff yields the original HDE model~\cite{Cohen1999, Hsu2004, Li2004}. This formulation, however, is unable to describe the complete history of cosmic evolution~\cite{Li2004, Wang2017}, failing to reproduce the correct dynamical behavior across the radiation dominated, matter dominated, and dark energy dominated epochs. To address this shortcoming, researchers have proposed various HDE models by considering different forms of horizon entropy or alternative IR cutoffs~\cite{Wang2017}. For instance, Tsallis HDE models are based on Tsallis entropy as a generalization of the standard Boltzmann--Gibbs entropy~\cite{Tavayef2018}, while Barrow HDE models originate from Barrow's modification of the Bekenstein--Hawking entropy formula incorporating quantum gravitational corrections with a fractal horizon structure~\cite{Saridakis2020}. Recently, motivated by fractional entropy that originates from fractional quantum mechanics in the context of Schwarzschild black hole thermodynamics~\cite{Jalalzadeh2021}, fractional holographic dark energy (FHDE) has been proposed~\cite{Trivedi2024}. This model successfully describes the late time acceleration of the universe for small values of the model parameter $\alpha$, but it is classically unstable since the squared sound speed is negative. In contrast, in the Tsallis and Barrow HDE models, the classical instability is removed once an interaction between the dark matter and dark energy sectors is considered~\cite{Huang2019, Huang2021}.

Observational constraints are crucial for assessing the viability of HDE models, necessitating comparison with precise data, including the Pantheon+ SN Ia sample~\cite{Riess2022, Brout2022, Scolnic2022}, observational Hubble parameter data (OHD)~\cite{Cao2022}, the Dark Energy Spectroscopic Instrument (DESI)~\cite{Abdul2025}, and the Cosmic Microwave Background (CMB)~\cite{Planck2020}. As standardizable candles, SN Ia allow for accurate distance measurements and were pivotal in revealing the late time acceleration of the universe. In parallel, OHD, inferred from galaxy ages via the cosmic chronometers approach, provide direct insights into the expansion history. Recently, the DESI collaboration released the baryon acoustic oscillation measurements from its second data release (DR2), which represent the largest spectroscopic galaxy sample to date and serve as a standard probe of the cosmic expansion history. As a relic radiation from the recombination epoch, the CMB encodes the primordial density fluctuations and provides a firm anchor for the cosmic distance scale, enabling robust tests of dark energy models when combined with low-redshift data. As a result, researchers routinely combine these datasets to test the consistency of dark energy models and place stringent constraints on their key parameters~\cite{Shen2026, Zhang2026, Huang2025Aa, Liu2023, Shen2025, Wang2025, Liu2024, Liu2024b, Arora2024, Oliveros2024, Wang2023, Mukherjee2022a, Mukherjee2022, Cao2021a, Pacif2021, Cao2021, Akarsu2020, Yang2020, Liu2019, Jimenez2016, Gong2013, Su2011, Gong2010, Wu2010a, Gong2006, Wu2006, Yadav2026, Zhu2026, Plaza2025, Petri2026, Li2026, Li2025, Li2025a, Li2025b, Luciano2026, Wang2026}. However, when different datasets are analyzed separately, notable discrepancies emerge. One is the Hubble tension, a significant disagreement between local $H_{0}$ measurements from SH0ES and the CMB inferred value. The expansion rate from the early universe CMB data is $H_{0} = 67.4 \pm 0.5 \mathrm{km\,s}^{-1}\,\mathrm{Mpc}^{-1}$~\cite{Planck2020}, while the late universe distance ladder gives $H_{0} = 73.04 \pm 1.04 \mathrm{km\,s}^{-1}\,\mathrm{Mpc}^{-1}$~\cite{Riess2022}. The other is the $\sigma_{8}$ ($S_{8}$) tension, where CMB based estimates of matter clustering systematically exceed those from weak gravitational lensing surveys. It is found that Hubble tension can be partially mitigated in HDE with the future event horizon as the IR cutoff~\cite{Li2026a}, whereas for the interacting dark energy, the Hubble tension can be alleviated~\cite{Silva2025}; for Barrow interacting HDE with the Hubble horizon as the IR cutoff, both the Hubble tension and $S_{8}$ tension can be alleviated~\cite{Yarahmadi2025}. It is therefore natural to ask whether the interacting fractional holographic dark energy (IFHDE) model can be supported by observational data and whether it has the potential to alleviate the Hubble tension and eliminate the classical instability by introducing an interaction term between the pressureless matter and FHDE.

A successful dark energy model must not only account for the late time acceleration of the universe under observational constraints but also describe the entire evolution of the universe, namely that the universe evolves from the radiation dominated epoch into the matter dominated epoch and eventually enters the dark energy dominated epoch. The phase space analysis provides an effective approach to studying this evolutionary process. In this method, the critical points of a dynamical system characterize the key phases of cosmic evolution, with a stable fixed point representing the dark energy dominated epoch~\cite{Bahamonde2018}. This approach has also been successfully applied to HDE models, yielding significant results~\cite{Setare2009, Liu2010, Banerjee2015, Mahata2015, Mishra2019, Bargach2019, Tita2024}, including Tsallis holographic dark energy~\cite{Huang2019, Ebrahimi2020, Astashenok2023} and Barrow holographic dark energy~\cite{Huang2021, Srivastava2021}. Therefore, it is worthwhile to investigate whether the IFHDE model can describe the whole evolution of the universe and whether it can be distinguished from the $\Lambda$CDM model.

The goals of this paper are twofold: to constrain the parameters of the IFHDE model and to examine its viability in describing the universe's entire evolutionary history. This paper is structured as follows. In Section~\ref{sec:2}, we introduce the IFHDE model. Section~\ref{sec:3} constrains the parameter of the IFHDE model. Section~\ref{sec:4} examines its viability in describing the universe's entire evolutionary history by the phase space analysis and statefinder diagnostic. Our main conclusions are summarized in Section~\ref{sec:5}.

\section{Model} \label{sec:2}

With the fractional entropy and the Hubble horizon as the IR cutoff, the FHDE density is given by~\cite{Trivedi2024}
\begin{equation}
\rho_{de}=3 c^{2} L^{\frac{2-3\alpha}{\alpha}}=3 c^{2} H^{\frac{3\alpha-2}{\alpha}},
\end{equation}
where $\alpha$ is a fractional parameter constrained to $1 < \alpha \leq 2$. For $\alpha=2$, this model reduces to the standard HDE. It also encompasses the Barrow and Tsallis HDE models when $\alpha$ takes the forms $\alpha=\frac{2}{\Delta+1}$ and $\alpha=\frac{2}{2\delta-1}$, respectively.

We consider a homogeneous and isotropic Friedmann--Robertson--Walker universe described by the line element
\begin{equation}
ds^{2}=-dt^{2}+a^{2}(t)(dr^{2}+r^{2}d\Omega^{2}),
\end{equation}
the Friedmann equation is
\begin{equation}
H^{2}=\frac{\kappa^{2}}{3}\big(\rho_{r}+\rho_{m}+\rho_{de}\big),\label{H2}
\end{equation}
where $\rho_{r}$, $\rho_{m}$, and $\rho_{de}$ denote the energy densities of radiation, pressureless matter, and FHDE, respectively, and their conservation equations are
\begin{eqnarray}
&& \dot{\rho}_{r}+3H\rho_{r}=0,\label{rhor}\\
&& \dot{\rho}_{m}+3H\rho_{m}=Q,\label{rhom}\\
&& \dot{\rho}_{de}+3H(1+\omega_{de})\rho_{de}=-Q,\label{rhode}
\end{eqnarray}
with the equation of state parameter defined as
\begin{equation}
\omega_{de}=\frac{p_{de}}{\rho_{de}}.
\end{equation}

Here, 
$Q$ characterizes the energy exchange between the FHDE and pressureless matter sectors. For $Q>0$, energy transfers from FHDE to pressureless matter; for $Q<0$, the direction is reversed, with energy flowing from pressureless matter to FHDE.

To analyze the dynamical evolution of the universe, we introduce the following dimensionless variables
\begin{equation}
\Omega_{r}=\frac{\kappa^{2}\rho_{r}}{3H^{2}}, \quad \Omega_{m}=\frac{\kappa^{2}\rho_{m}}{3H^{2}}, \quad \Omega_{de}=\frac{\kappa^{2}\rho_{de}}{3H^{2}}, \quad \sigma=\frac{\kappa^{2}Q}{3H^{2}},
\end{equation}
which allow us to rewrite the Friedmann Equation~(\ref{H2}) as
\begin{equation}
\Omega_{r}+\Omega_{m}+\Omega_{de}=1.\label{O1}
\end{equation}

Combining Equations~(\ref{H2})--(\ref{rhode}) and~(\ref{O1}) yields
\begin{equation}
\frac{\dot{H}}{H^{2}}=\frac{1}{2}\Big[\Omega_{m}+(1-3\omega_{de})\Omega_{de}\Big]-2.\label{HH2}
\end{equation}

The deceleration parameter $q$ is then given by
\begin{equation}
q=-1-\frac{\dot{H}}{H^{2}},\label{q0}
\end{equation}
and the squared sound speed $v^{2}_{s}$ is defined by
\begin{equation}
v^{2}_{s}=\frac{\rho_{de}}{\dot{\rho}_{de}}\dot{\omega}_{de}+\omega_{de},\label{vs20}
\end{equation}
The sign of the squared sound speed $v^{2}_{s}$ thus serves as a stability criterion, with the model being stable for $v^{2}_{s}>0$ and unstable otherwise.

To identify a viable IFHDE model, following the widely adopted parametrizations of the interacting dark energy, we examine three forms with the following interaction terms~\cite{Bahamonde2018, Silva2025, Wang2024}: 

\begin{enumerate}
\item[(i)] IFHDE-A: $Q=\beta H \rho_{m}$;
\item[(ii)] IFHDE-B: $Q=\gamma H \rho_{de}$;
\item[(iii)] IFHDE-C: $Q=\beta H \rho_{m}+\gamma H \rho_{de}$.
\end{enumerate}
The dimensionless parameters $\beta$ and $\gamma$ determine the strength of the interaction, with their signs indicating the direction of energy transfer. Specifically, $\beta > 0$ or $\gamma > 0$ corresponds to energy transfer from dark energy to dark matter, while negative values correspond to the opposite direction. The specific form $Q$ is purely phenomenological~\cite{Silva2025}, motivated by dimensional consistency, the minimality of free parameters, and direct comparability with existing observational constraints. While the original motivation for dark sector interactions was to address the coincidence problem, recent emphasis has shifted towards alleviating the Hubble tension, i.e., the discrepancy between CMB-derived and local measurements of the Hubble constant.

Using Equations~(\ref{rhom}),~(\ref{rhode}),~(\ref{O1}) and~(\ref{HH2}), together with the definition $'=\frac{d}{d(\ln a)}$, we derive the automatic dynamical equations governing this system
\begin{eqnarray}
&& \Omega'_{m}=[(3\omega_{de}-1)\Omega_{de}-\Omega_{m}+1]\Omega_{m}+\sigma,\label{Omm}\\
&& \Omega'_{de}=[(3\omega_{de}-1)(\Omega_{de}-1)-\Omega_{m}]\Omega_{de}-\sigma,\label{Omde}\\
\end{eqnarray}
with
\begin{equation}
\omega_{de}=\frac{[(3\alpha-2)(\Omega_{m}+\Omega_{de})+8-6\alpha]\Omega_{de}+2\alpha\sigma}{3[(3\alpha-2)\Omega_{de}-2\alpha]\Omega_{de}}.\label{w0}
\end{equation}
Under the conditions $\sigma=0$ and $\Omega_{m}=1-\Omega_{de}$, Equation~(\ref{w0}) reduces to the standard FHDE model~\cite{Trivedi2024}. In FHDE, a small value of $\alpha$ is required to obtain an acceptable cosmic evolution, with the equation of state parameter $\omega_{de}$ at $\alpha=1.1$ being in close agreement with the recent DESI constraints~\cite{Trivedi2024}.
 
\section{Observational Constraints} \label{sec:3}
 
To estimate the parameters for the IFHDE models, we combine the latest observational data including SN Ia, OHD, BAO, and CMB. To achieve this goal, we use the public code EMCEE in Python for implementing the Markov Chain Monte Carlo (MCMC) sampling. For the IFHDE-A model, we constrain six parameters $\{ H_{0}, \Omega_{m,0}, \Omega_{de,0}, \alpha, \beta, M \}$; for the IFHDE-B model, we constrain $\{ H_{0}, \Omega_{m,0}, \Omega_{de,0}, \alpha, \gamma, M \}$; for the IFHDE-C model, we constrain $\{ H_{0}, \Omega_{m,0}, \Omega_{de,0}, \alpha, \beta, \gamma, M \}$.

For the IFHDE models, the expansion rate function $E(z)$ can be expressed as
\begin{equation}
E(z)=\frac{H(z)}{H_{0}}=\sqrt{\Omega_{r,0}(1+z)^{4}+\Omega_{m,0}(1+z)^{3}+\Omega_{de,0}e^{3\int_{0}^{z}\frac{1+\omega_{de}}{1+z}dz}},
\end{equation}
where $\Omega_{r,0}+\Omega_{m,0}+\Omega_{de,0}=1$.

The $\chi^{2}$ statistic is adopted to evaluate the consistency between theoretical predictions and observational data. Minimizing this statistic allows us to identify the set of model parameters that provides the best description of the observed universe.

\subsection{Type Ia Supernovae}

For SN Ia data, we use Pantheon+ sample covering the redshift range $z \in [0.01, 2.261]$~\cite{Riess2022, Brout2022, Scolnic2022}. The theoretical apparent magnitude $m_{th}$, representing the predicted observable for SN Ia, is expressed as
\begin{equation}
m_{th}=5 \log_{10} \Big( \frac{D_{L}(z)}{Mpc} \Big) + 25 + M,
\end{equation}
where $M$ is the absolute magnitude of SN Ia, the luminosity distance $D_{L}(z)$ is given by
\begin{equation}
D_{L}(z)=c(1+z)\int^{z}_{0}\frac{dz}{H(z)},
\end{equation}
and the $\chi^{2}_{SN}$ term for the SN Ia sample is defined as
\begin{equation}
\chi^{2}_{SN}=\left( \hat{m}_{obs} - m_{th} \right)^{T} C_{SN}^{-1} \left( \hat{m}_{obs} - m_{th} \right),
\end{equation}
where $\hat{m}_{obs}$ is the array of observed corrected apparent magnitude, and $C_{SN}$ denotes the associated covariance matrix.

\subsection{Hubble Parameter Data}

For the OHD data, we adopt the data spanning $z \in [0.07, 1.965]$ collected by~\cite{Cao2022}, with the corresponding $\chi^{2}_{OHD}$ term evaluated as
\begin{equation}
\chi^{2}_{OHD} = \sum_{i=1}^{N_{OHD}} \left(\frac{H_{obs,i}-H_{th}(z_i)}{\sigma_{OHD,i}}\right)^2
\end{equation}
where $H_{obs,i}$ and $\sigma_{OHD,i}$ represent the $i$-th observed value and its associated standard deviation, respectively, and $N_{OHD}$ denotes the total number of OHD data points.

\subsection{Baryon Acoustic Oscillation}

For the BAO data, we use the latest DESI DR2 BAO data covering $z \in [0.295,2.33]$~\cite{Abdul2025}, with the $\chi^{2}_{DESI}$ term defined as
\begin{equation}
\chi^{2}_{DESI} = \sum_{i=1} \Delta_{i}^{T} \mathbf{C}_{i}^{-1} \Delta_{i},
\end{equation}
with
\begin{eqnarray}
\Delta_{i}=
\left(
\begin{matrix}
D_{M}^{obs}(z_{i})/r_{d}-D_{M}^{th}(z_{i})/r_{d}\\[6pt]
D_{H}^{obs}(z_{i})/r_{d}-D_{H}^{th}(z_{i})/r_{d}
\end{matrix}
\right)
\end{eqnarray}
and 
\begin{eqnarray}
\mathbf{C}_{i}=
\left(
\begin{matrix}
\sigma^{2}_{D_M/r_d} & r_{HM} \sigma_{D_M/r_d} \sigma_{D_H/r_d}\\
r_{HM} \sigma_{D_M/r_d} \sigma_{D_H/r_d} & \sigma^{2}_{D_H/r_d}
\end{matrix}
\right)
\end{eqnarray}
in which the Hubble distance and the transverse comoving distance are given by
\begin{equation}
D_{H}(z)=\frac{c}{H(z)},
\end{equation}
and
\begin{equation}
D_{M}(z)=c\int^{z}_{0}\frac{dz}{H(z)},
\end{equation}
where $\sigma_{D_{M}/r_{d}}$ and $\sigma_{D_{H}/r_{d}}$ represent the observational uncertainties associated with $D_{M}(z)/r_{d}$ and $D_{H}(z)/r_{d}$, respectively; $r_{HM}$ denotes their correlation coefficient, and $r_{d}$ is the sound horizon at the drag epoch.

\subsection{Cosmic Microwave Background}

For the CMB data, we use the distance priors of Planck 2018 from~\cite{Zhai2019,Chen2019}, a method that compresses the full CMB data into background quantities, allowing the full CMB power spectrum to be substituted with a more compact representation while preserving key cosmological information~\cite{Efstathiou1999, Wang2006, Wang2007}. The $\chi^{2}_{CMB}$ term can be written as
\begin{equation}
\chi^{2}_{CMB} = \Delta p^{T} \mathbf{C}_{CMB}^{-1} \Delta p, \qquad \Delta p=p^{obs}-p^{th},
\end{equation}
where $p=\{R,l_{A},\Omega_{b}h^{2}\}$, and $\mathbf{C}_{CMB}$ is the covariance matrix. The CMB shift parameter $R$ and the acoustic scale $l_{A}$ are defined as
\begin{equation}
R=\frac{D_{M}(z_{*})\sqrt{\Omega_{m}H^{2}_{0}}}{c}, \qquad l_{A}=\frac{\pi D_{M}(z_{*})}{r_{s}(z_{*})},
\end{equation}
where $z_{*}$ denotes the redshift corresponding to the photon decoupling epoch. The Planck CMB observations give $p^{\text{obs}}=\{1.7502, 301.471, 0.02236\}$. Since $\Omega_{b}h^{2}$ is correlated with $R$ and $l_{A}$, their covariance matrix, adopted from~\cite{Chen2019}, is also included in the $\chi^{2}$ calculation.

\subsection{Results}

The total log-likelihood is defined as
\begin{equation}
\ln(\mathcal{L}_{total})=-\frac{1}{2}\chi^{2}_{total}+const.,
\end{equation}
with
\begin{equation}
\chi^{2}_{total}=\chi^{2}_{SN}+\chi^{2}_{OHD}+\chi^{2}_{DESI}+\chi^{2}_{CMB}.
\end{equation}
To enable a statistical comparison between the IFHDE models with the standard $\Lambda$CDM model, we also perform a corresponding MCMC analysis of $\Lambda$CDM using the identical dataset. Given the disparity in the numbers of free parameters, we apply the Akaike Information Criterion (AIC)~\cite{Akaike1974} for model selection, defined {as} 
\begin{equation}
AIC=\chi^{2}_{min}+2n,
\end{equation}
where $n$ is the number of model parameters. Since only the relative differences of AIC 
between models are meaningful, we consider $\Delta$AIC, defined as $\Delta \mathrm{AIC}=\mathrm{AIC_{model}}-\mathrm{AIC}_{\Lambda \mathrm{CDM}}$. Compared with the reference model, $\Delta \mathrm{AIC} < 0$ favors the model; $0<\Delta \mathrm{AIC} < 2$ indicates substantial support; $4 < \Delta \mathrm{AIC} < 7$ indicates considerably weaker support; and $\Delta \mathrm{AIC} > 10$ indicates essentially no support.
 
\begin{figure}[H]
\begin{center}
\includegraphics[width=0.8\textwidth]{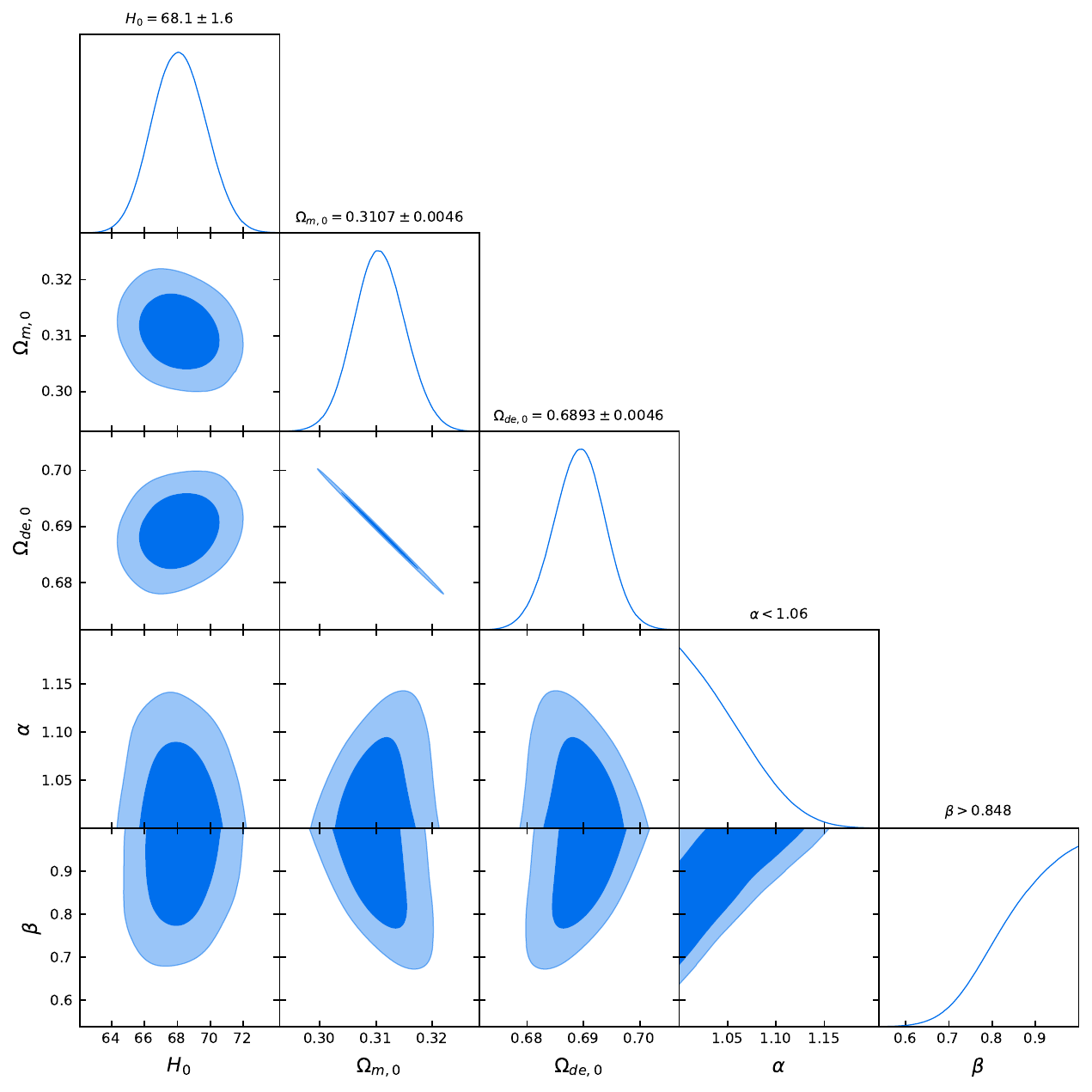}
\caption{\label{Fig1}Confidence contours for the model parameters of the IFHDE-A model using SNIa, OHD, DESI DR2, and CMB datasets.}
\end{center}
\end{figure}

\begin{figure}[H]
\begin{center}
\includegraphics[width=0.81\textwidth]{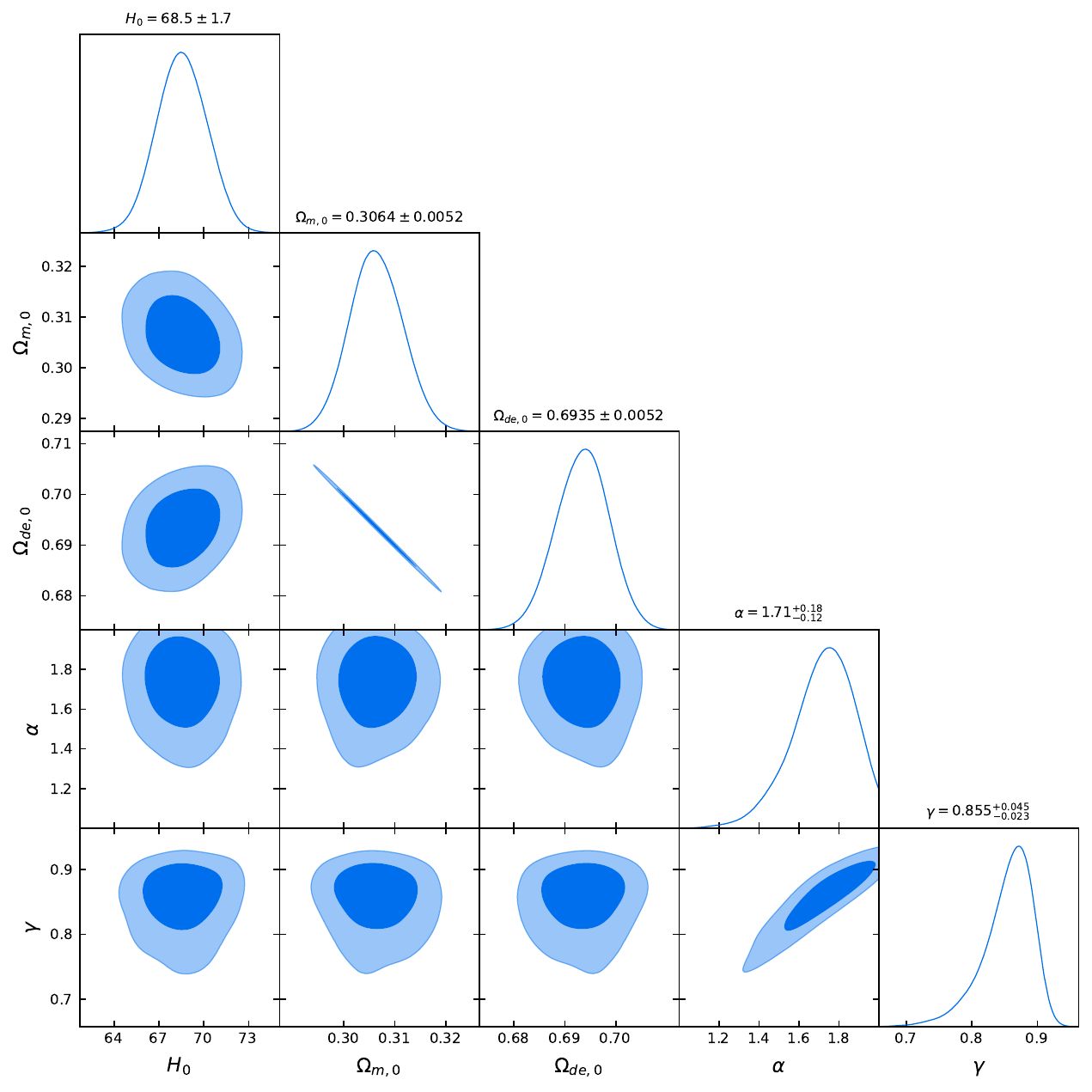}
\caption{\label{Fig2}Confidence contours for the model parameters of the IFHDE-B model using SNIa, OHD, DESI DR2, and CMB datasets.}
\end{center}
\end{figure}

\begin{figure}[H]
\begin{center}
\includegraphics[width=0.81\textwidth]{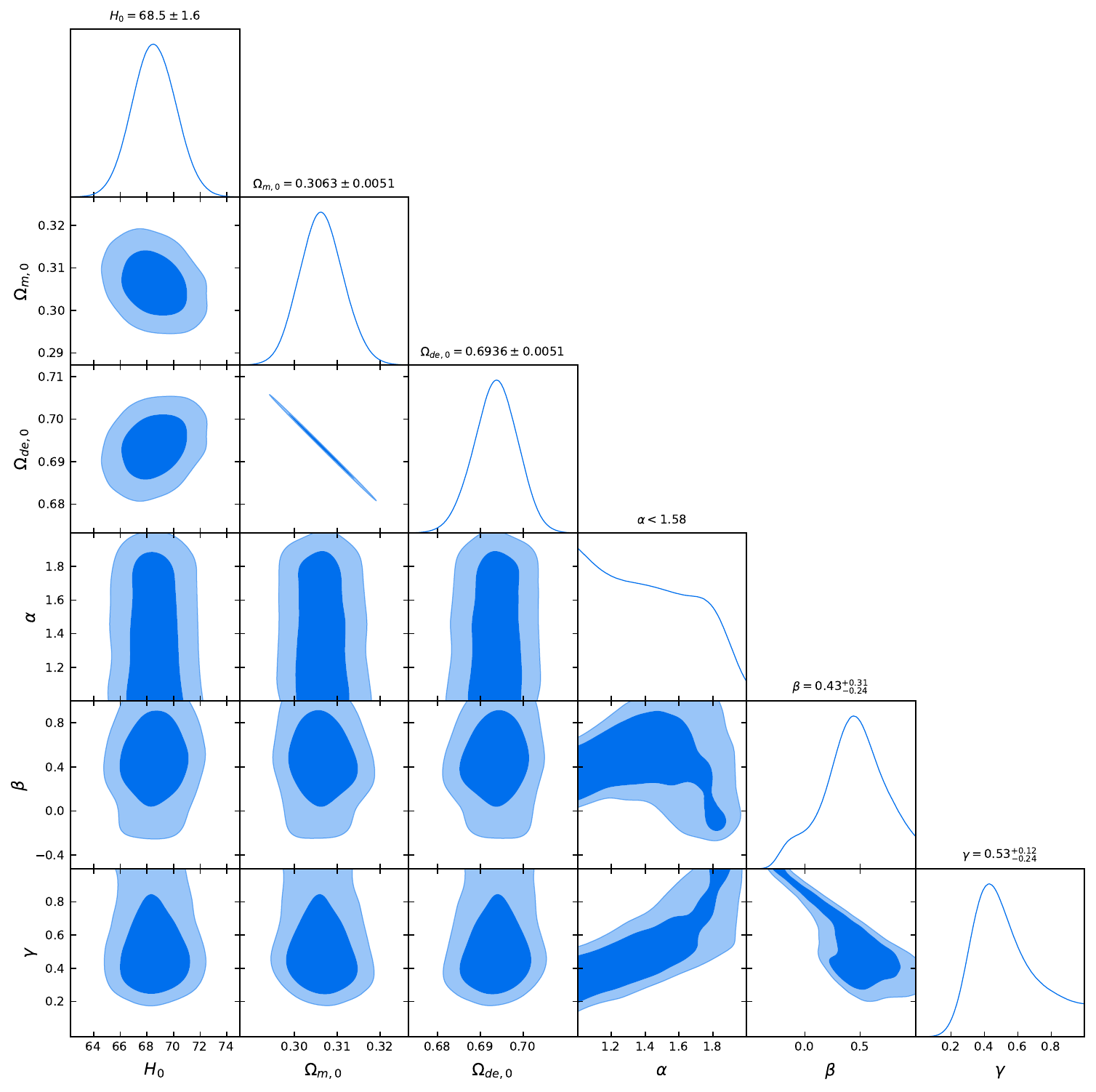}
\caption{\label{Fig3}Confidence contours for the model parameters of the IFHDE-C model using SNIa, OHD, DESI DR2, and CMB datasets.}
\end{center}
\end{figure}

We summarize the parameter constraints for the $\Lambda$CDM, IFHDE-A, IFHDE-B, and IFHDE-C models in Table~\ref{Tab1}, where the mean values and 1$\sigma$ confidence levels (CL) are listed. The posterior distributions for the IFHDE-A, IFHDE-B, and IFHDE-C models are presented in Figures~\ref{Fig1},~\ref{Fig2} and~\ref{Fig3}, respectively. The absolute magnitude $M$ of SN Ia is included as a free parameter in all models, but its constrained values are omitted here for brevity.

\begin{table}[H]
\renewcommand{\arraystretch}{1.5} 
\centering
\caption{\label{Tab1}Observational constraints for $\Lambda$CDM, IFHDE-A, IFHDE-B, and IFHDE-C models.}
 \begin{tabular}{ccccc}
\toprule
  \textbf{Parameters} & \boldmath\textbf{$\Lambda$CDM} & \textbf{IFHDE-A} & \textbf{IFHDE-B} & \textbf{IFHDE-C}\\
  \midrule
  $H_{0}$ & $68.8 \pm 1.6$ & $68.1 \pm 1.6$ & $68.5 \pm 1.7$ & $68.5 \pm 1.6$\\
 
  $\Omega_{m,0}$ & $0.3046 \pm 0.0033$ & $0.3107 \pm 0.0046$ & $0.3064 \pm 0.0052$ & $0.3063 \pm 0.0051$\\    
 
  $\Omega_{de,0}$ & $0.6953 \pm 0.0033$ & $0.6893 \pm 0.0046$ & $0.6935 \pm 0.0052$ & $0.6936 \pm 0.0051$\\  
 
  $\alpha$ & $-$ & $<$1.06 & $1.71^{+0.18}_{-0.12}$ & $<$1.58\\
 
  $\beta$ & $-$ & $>$0.848 & $-$ & $0.43^{+0.31}_{-0.24}$\\
 
  $\gamma$ & $-$ & $-$ & $0.855^{+0.045}_{-0.023}$ & $0.53^{+0.12}_{-0.24}$\\
 
  $\chi^{2}_{min}$ & $1437.4$ & $1448.9$ & $1434.7$ & $1432.9$\\
 
  $\Delta \chi^{2}_{min}$ & $0$ & $11.5$ & $-2.7$ & $-4.5$\\
 
  $AIC$ & $1445.4$ & $1460.9$ & $1446.4$ & $1446.9$\\
 
  $\Delta AIC$ & $0$ & $15.5$ & $1$ & $1.5$\\
\bottomrule
  \end{tabular}
\end{table}

As shown in Table~\ref{Tab1}, the IFHDE-A, IFHDE-B, and IFHDE-C models yield $H_{0}$ values consistent with Planck 2018 within $1\sigma$, with IFHDE-A giving $68.1 \pm 1.6$, IFHDE-B giving $68.5 \pm 1.7$, and IFHDE-C giving $68.5 \pm 1.6$. None of the models approaches the local measurement of $73.04 \pm 1.04 \mathrm{km\,s}^{-1}\,\mathrm{Mpc}^{-1}$ from SH0ES, indicating that the IFHDE-A, IFHDE-B, and IFHDE-C models fail to alleviate the Hubble tension.

Based on the results in Table~\ref{Tab1}, among the three IFHDE models, IFHDE-C yields the smallest $\chi^{2}_{min}$ value, implying that it provides the best fit to the data, comparable to the $\Lambda$CDM model. IFHDE-B also yields a $\chi^{2}_{min}$ value lower than that of $\Lambda$CDM model. In contrast, IFHDE-A gives the largest $\chi^{2}_{min}$, indicating the poorest fit. Because $\chi^{2}_{min}$ does not account for the number of free parameters, we adopt the AIC to enable a more stringent model comparison. Among these models, $\Lambda$CDM yields the smallest AIC, indicating a better balance between goodness of fit and model complexity under this criterion, whereas IFHDE-A yields the largest AIC.

\vspace{-12pt}

Compared to the $\Lambda$CDM model, IFHDE-A gives $\Delta \chi^{2}_{min}=11.5$ and $\Delta \mathrm{AIC}=15.5$, indicating that it is not supported by the data. In contrast, both IFHDE-B and IFHDE-C show some preference 
by the AIC with $\Delta \mathrm{AIC}=1$ and $\Delta \mathrm{AIC}=1.5$, respectively. Notably, these two models achieve $\Delta \chi^{2}_{min}=-2.7$ and $\Delta \chi^{2}_{min}=-4.5$, meaning they fit the observational data as well as the $\Lambda$CDM model.

The 
OHD data employed here are subject to important limitations, as they are not universally accepted as a robust cosmological probe, due to issues related to age uncertainty estimation and model dependence~\cite{Kjerrgren2023}. Furthermore, the AIC used here provides only a rough indicative comparison and is not sufficient for definitive model selection or rigorous statistical inference. Therefore, our AIC results are presented as a preliminary guide and should not be overinterpreted as firm conclusions in favor of any particular model.

\section{Evolution of the Universe} \label{sec:4}

In the previous section, we constrained the model parameters of the IFHDE-A, IFHDE-B, and IFHDE-C models using the SN Ia, OHD, DESI DR2, and CMB datasets. The results indicate that the IFHDE-B and IFHDE-C are substantially supported by the  AIC. In this subsection, we analyze the cosmological evolution of these models. Due to the skewed posterior distributions, the mean values of the IFHDE-C model parameters suffer from large uncertainties. For a consistent comparison, we therefore adopt the best-fit values of both the IFHDE-B and IFHDE-C models to analyze the evolution of the universe. The best-fit values of the IFHDE-B and IFHDE-C models are $\{ H_{0}, \Omega_{m,0}, \Omega_{de,0}, \alpha, \gamma\}=\{ 68.263, 0.307, 0.693, 1.735, 0.861 \}$ and $\{ H_{0}, \Omega_{m,0}, \Omega_{de,0}, \alpha, \beta, \gamma \}=\{ 68.415, 0.308, 0.691, 1.042, 0.334, 0.363 \}$, respectively.

\subsection{Phase Space Analysis}

To analyze the evolution of the universe in the IFHDE-B and IFHDE-C models, we employ phase space analysis to investigate whether these models can reproduce the complete cosmic history, originating from the radiation dominated epoch, transitioning to the pressureless matter dominated epoch, and eventually entering the dark energy \mbox{dominated epoch.}

To achieve this goal, following Ref.~\cite{Bahamonde2018, Huang2019, Huang2021, Wu2010, Dutta2017, Wu2007, Wu2008, Huang2025a, Huang2025b}, we obtain the critical points by solving the corresponding autonomous system
\begin{equation}
\Omega'_{m}=\Omega'_{de}=0.\label{OO0}
\end{equation}
After solving Equation~(\ref{OO0}), we find that both the IFHDE-B and IFHDE-C models have three critical points, which are listed in Tables~\ref{Tab2} and~\ref{Tab3}, respectively. Then, linearizing the autonomous system, we can obtain a set of first order differential equations for this system. The stability of the critical points is then governed by the eigenvalues of the coefficient matrix of this linearized system. According to linear stability theory, a point is an attractor when all eigenvalues are negative, unstable when all are positive, and a saddle point when they have mixed signs. For the IFHDE-B and IFHDE-C models, the stability of their critical points is also summarized in Tables~\ref{Tab2} and~\ref{Tab3}.

\begin{table}[H]
\renewcommand{\arraystretch}{1.4} 
\caption{\label{Tab2}Critical points and their stability conditions for the IFHDE-B model.}
\centering
\small
\begin{tabular}{cccccccc}
\toprule
  \textbf{Label} & \boldmath\textbf{$(\Omega_{m},\Omega_{de})$} & \boldmath\textbf{$\Omega_{r}$} & \boldmath\textbf{$\omega_{de}$} & \boldmath\textbf{$q$} & \textbf{Eigenvalues} & \textbf{Conditions} & \textbf{Points}\\
  \midrule
  $B_{1}$ & $(0,0)$ & $1$ & $1-\frac{4}{3\alpha}-\frac{\gamma}{3}$ & $1$ & $(-2+\frac{4}{\alpha},1)$ & $1<\alpha<2$ & $Unstable$\\
 
  $B_{2}$ & $(1,0)$ & $0$ & $\frac{1}{2}-\frac{1}{\alpha}-\frac{\gamma}{3}$ & $\frac{1}{2}$ & $(-\frac{3}{2}+\frac{3}{\alpha},-1)$ & $1<\alpha<2$ & $Saddle$\\
 
  $B_{3}$ & $(\frac{\gamma}{3+\gamma},\frac{3}{3+\gamma})$ & $0$ & $-1-\frac{\gamma}{3}$ & $-1$ & $(\frac{3(\alpha-2)(3+\gamma)}{6-\alpha(3-2\gamma)},-4)$ & $1<\alpha<2,0\leq\gamma\leq1$ & $Stable$\\
 \bottomrule
  \end{tabular}
\end{table}
\vspace{-9pt}
\begin{table}[H]
\renewcommand{\arraystretch}{1.4} 
\caption{\label{Tab3}Critical points and their stability conditions for the IFHDE-C model.}
\begin{adjustwidth}{-\extralength}{-\extralength}
\centering
\small
\begin{tabular}{cccccccc}
\toprule
  \textbf{Label} & \boldmath\textbf{$(\Omega_{m},\Omega_{de})$} & \boldmath\textbf{$\Omega_{r}$} & \boldmath\textbf{$\omega_{de}$ }&\boldmath\textbf{ $q$} & \textbf{Eigenvalues} & \textbf{Conditions} & \textbf{Points}\\
\midrule
  $C_{1}$ & $(0,0)$ & $1$ & $1-\frac{4}{3\alpha}-\frac{\gamma}{3}$ & $1$ & $(-2+\frac{4}{\alpha},1+\beta)$ & $1<\alpha<2,0\leq\beta\leq1$ & $Unstable$\\
 
  $C_{2}$ & $(1,0)$ & $0$ & $-$ & $\frac{1}{2}$ & $(\frac{(2-\alpha)(3-\beta)}{2\alpha},-1-\beta)$ & $1<\alpha<2,0\leq\beta\leq1$ & $Saddle$\\
 
  $C_{3}$ & $(\frac{\gamma}{3+\gamma-\beta},\frac{3-\beta}{3+\gamma-\beta})$ & $0$ & $-1-\frac{\gamma}{3-\beta}$ & $-1$ & $(-\frac{(2-\alpha)(3-\beta)(3-\beta+\gamma)}{(2-\alpha)(3-\beta)+2\alpha\gamma},-4)$ & $1<\alpha<2,0\leq\beta\leq1,0\leq\gamma\leq1$ & $Stable$\\
 \bottomrule
  \end{tabular}
\end{adjustwidth}
\end{table}

For the IFHDE-B model, as shown in Table~\ref{Tab2}, points $B_{1}$ and $B_{2}$ represent decelerated epochs, as indicated by their positive $q$ values, while point $B_{3}$ denotes an accelerated epoch. For all points, the equation of state parameter $\omega_{de}$ is determined by $\alpha$ and $\gamma$. Based on the values of $\Omega_{m}$, $\Omega_{de}$, and $\Omega_{r}$, the three points can be categorized as follows: point $B_{1}$ corresponds to the radiation dominated deceleration epoch, point $B_{2}$ represents the pressureless matter dominated deceleration epoch, and point $B_{3}$ denotes the dark energy dominated acceleration epoch. The eigenvalues of these points indicate that $B_{1}$ is unstable, $B_{2}$ is a saddle point, and $B_{3}$ is an attractor. 

For the IFHDE-C model, the results summarized in Table~\ref{Tab3} indicate that point $C_{1}$ represents an unstable radiation-dominated deceleration epoch, $C_{2}$ corresponds to a pressureless matter-dominated deceleration epoch that is a saddle point, and $C_{3}$ denotes a dark energy-dominated acceleration epoch that is an attractor. The coordinate of point $C_{3}$, the equation of state parameter $\omega_{de}$, and the eigenvalues of these points are determined by $\alpha$, $\beta$, and $\gamma$. For point $C_{2}$, the value of $\omega_{de}$ does not exist; however, since this point is a saddle point, the evolution curves of the universe will not pass through it.

For point $C_{3}$, the coordinates and $\omega_{de}$ are determined by $\beta$ and $\gamma$, while those of point $B_{3}$ are determined by $\gamma$ alone. Thus, these points cannot behave as the cosmological constant, except when $\gamma = 0$.

\begin{figure}[H]
\begin{center}
\includegraphics[width=0.45\textwidth]{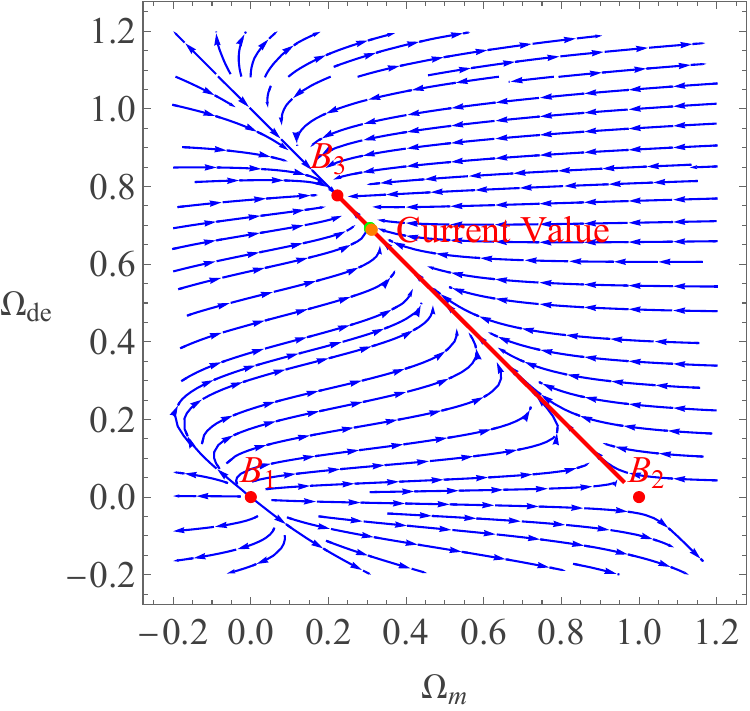}
\includegraphics[width=0.45\textwidth]{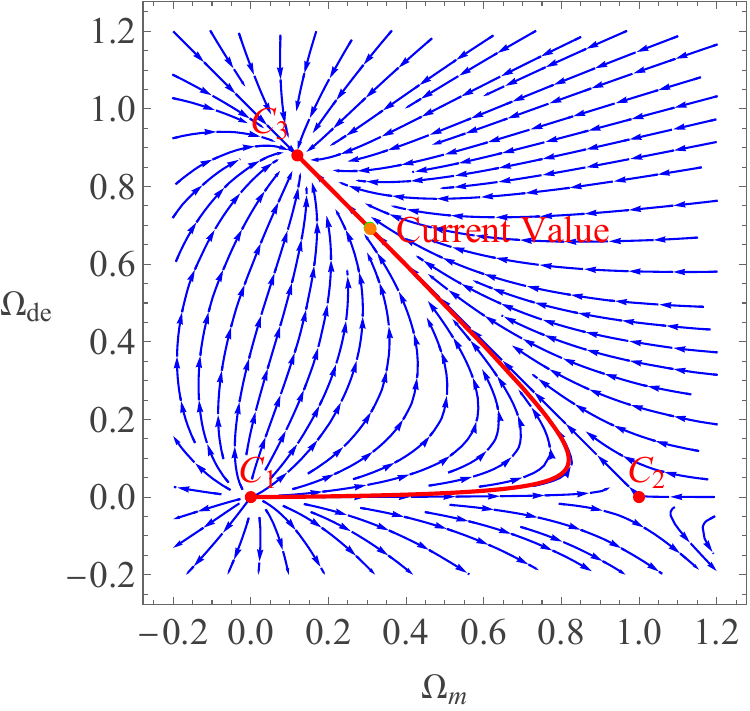}
\caption{\label{Fig4}Phase 
 space trajectories. The left panel is plotted for IFHDE-B with the best-fit values, while the right one is for IFHDE-C.}
\end{center}
\end{figure}

According to the stability of the critical points for the IFHDE-B and IFHDE-C models, the universe can evolve from the radiation-dominated epoch ($B_{1}/C_{1}$) through the matter-dominated epoch ($B_{2}/C_{2}$) into the dark energy-dominated late time acceleration epoch ($B_{3}/C_{3}$). In order to check whether these models can describe the entire evolution epoch of the universe when their parameters are set to the best-fit values, we plot the phase space trajectories of these models, as shown in Figure~\ref{Fig4}. In this figure, the green point denotes the current value from the best-fit values, while the orange one corresponds to the current value from the Planck 2018 results; the left panel shows IFHDE-B with the best-fit values, while the right panel shows IFHDE-C. 

The left panel of Figure~\ref{Fig4} shows that, for the IFHDE-B model, the universe evolves from the pressureless matter-dominated epoch into the dark energy-dominated epoch, rather than from the radiation-dominated epoch. Therefore, the IFHDE-B model fails to describe the full evolution history of the universe. The right panel of Figure~\ref{Fig4} shows that, for the IFHDE-C model, the universe stems from the radiation-dominated epoch, then passes through the pressureless matter-dominated epoch into the dark energy-dominated epoch. Thus, the IFHDE-C model can achieve this, but the dark energy itself cannot mimic the cosmological constant at the late time acceleration epoch.

\subsection{Statefinder Diagnostic}

In the previous subsection, we adopted the phase space analysis method to discuss the evolution of the universe in the IFHDE-B and IFHDE-C models and found that the IFHDE-C model can describe the full evolution history of the universe, but the dark energy itself cannot mimic the cosmological constant. In this subsection, we employ the statefinder to analyze the difference between the IFHDE-C and $\Lambda$CDM models. The statefinder parameters $r$ and $s$ are geometrical diagnostics that depend only on the scale factor $a$ and are defined as~\cite{Sahni2003, Wu2005}
\begin{equation}
r=\frac{\dddot{a}}{a H^{3}}, \qquad s=\frac{r-1}{3(q-\frac{1}{2})},
\end{equation}
where $H$ is the Hubble parameter, and $q$ is the deceleration parameter defined in Equation~(\ref{q0}). By differentiating Equation~(\ref{HH2}), the statefinder parameters $r$ and $s$ are expressed in terms of $\Omega'_{m}$ and $\Omega'_{de}$. The evolution curves of the universe in the $r-s$ and $r-q$ parameter space are then obtained through numerical solution of Equations~(\ref{Omm}) and~(\ref{Omde}).

\begin{figure}[H]
\begin{center}
\includegraphics[width=0.46\textwidth]{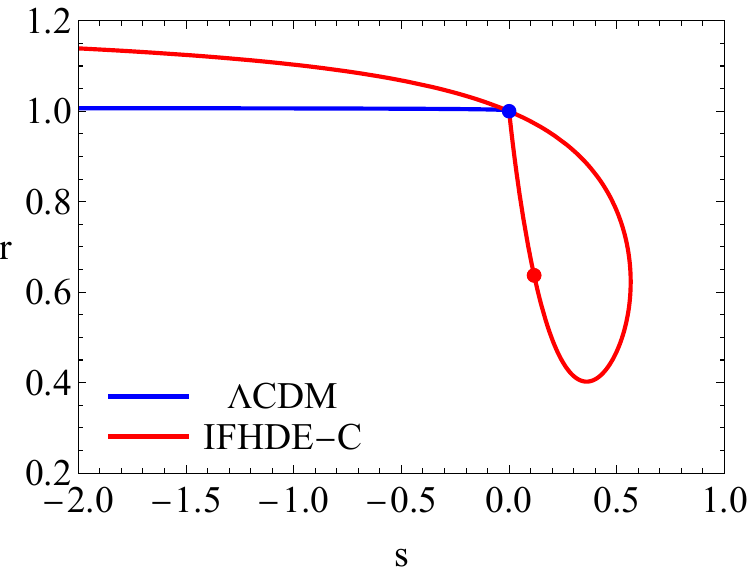}
\includegraphics[width=0.45\textwidth]{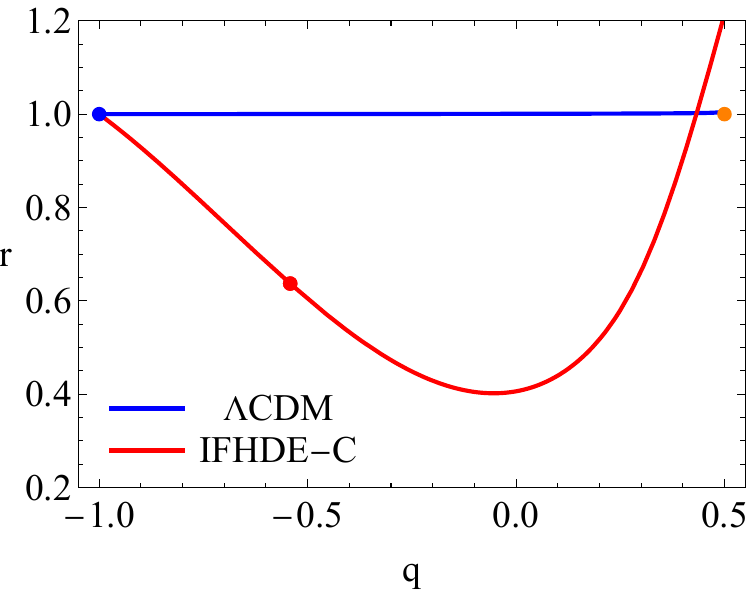}
\caption{\label{Fig5}Statefinder 
 diagnostics $\{r,s\}$ and $\{r,q\}$ for the IFHDE-C model with the best-fit values.}
\end{center}
\end{figure}

In the left panel of Figure~\ref{Fig5}, we plot the evolution curves of the statefinder diagnostic pair $\{s,r\}$; the blue dot marks the $\Lambda$CDM fixed point $(0,1)$, while the red dot represents the current value. This panel shows that the evolution curve for the IFHDE-C model deviates from $\Lambda$CDM but converges to the $\Lambda$CDM fixed point in the future. The right panel presents the evolution curves of the statefinder diagnostic pair $\{q,r\}$, where the blue dot marks the de Sitter expansion fixed point $(-1,1)$ in the future, the orange dot denotes the standard cold dark matter fixed point $(0.5,1)$, and the red dot represents the current value. This panel shows that the evolution curve for the IFHDE-C model deviates from $\Lambda$CDM but converges to the de Sitter expansion fixed point in the future.

Although the phase space analysis shows that the attractor $C_{3}$ of the IFHDE-C model has $\omega_{de} \neq -1$, the statefinder diagnostic of the IFHDE-C model reveals that the cosmic expansion converges to the $\Lambda$CDM fixed point $(s,r)=(0,1)$ and to the de Sitter expansion fixed point $(q,r)=(-1,1)$. This implies that, due to the interaction between dark matter and dark energy, the IFHDE-C model can mimic the $\Lambda$CDM at late times despite having a non-standard equation of state.

\subsection{Evolution of Cosmological Parameters}

In the previous subsection, we employed the statefinder diagnostic to compare the IFHDE-C model with the $\Lambda$CDM model and found that the statefinder pairs of the IFHDE-C model converge to the $\Lambda$CDM fixed point and to the de Sitter expansion fixed point. In this subsection, we further analyze the evolution of cosmological parameters for the IFHDE-C model to examine its late time cosmological behavior.

\begin{figure}[H]
\begin{center}
\includegraphics[width=0.45\textwidth]{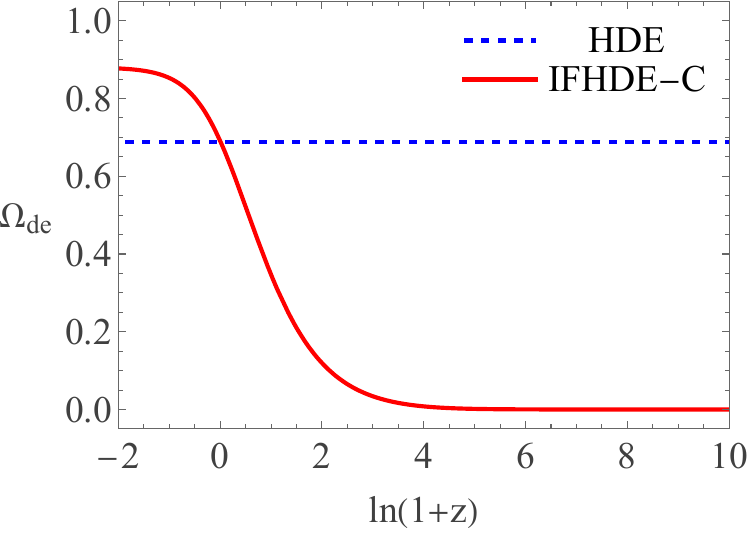}
\includegraphics[width=0.463\textwidth]{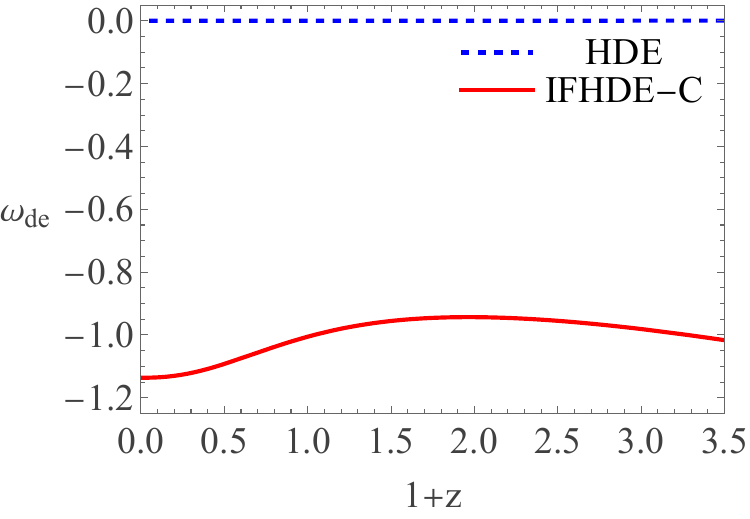}

\includegraphics[width=0.451\textwidth]{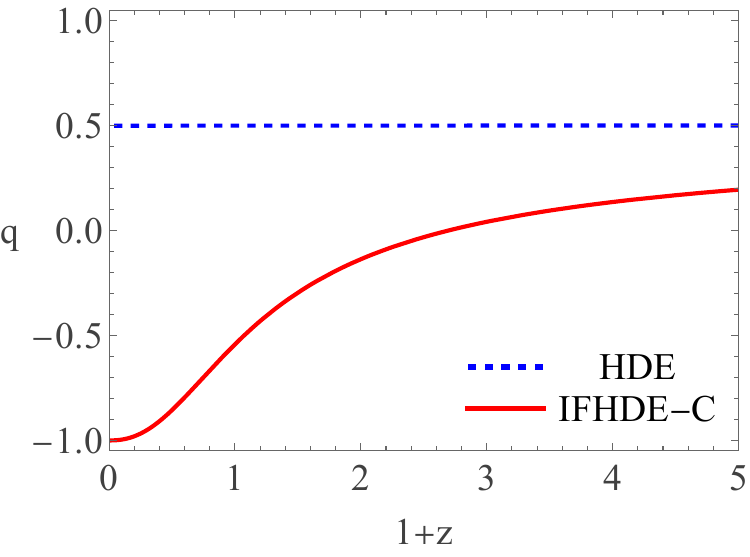}
\includegraphics[width=0.46\textwidth]{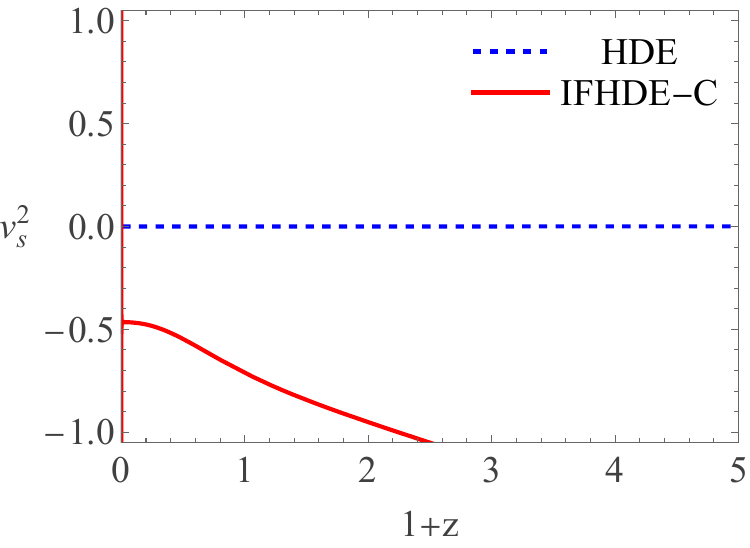}
\caption{\label{Fig6}Cosmological evolution and stability of the IFHDE-C model with the best-fit values.}
\end{center}
\end{figure}

By numerically solving Equations~(\ref{Omm}) and (\ref{Omde}) for the best-fit values, we obtain the evolutionary curves of $\Omega_{de}$, $\omega_{de}$, $q$, and $v^{2}_{s}$ as shown in Figure~\ref{Fig6}. In this figure, we add curves presenting the standard HDE plotted as a blue dashed line. The first panel shows that $\Omega_{de} \to 0$ at early times, while at late times it approaches $\Omega_{de} \to 0.88$, indicating that the late time evolution of the universe is dominated by dark energy. The second panel plots the evolution of $\omega_{de}$, which behaves like quintessence or phantom at the late times, while the third panel plots the evolution of $q$, where a transition from deceleration to acceleration occurs and $q$ also approaches $-1$ at late times. Both the second and third panels indicate that this model can realize late time acceleration. The fourth panel demonstrates that the IFHDE-C model exhibits classical instabilities under the best-fit parameters, as indicated by the negative values of $v_s^2$.

Since the IFHDE-C model is classically unstable with the best-fit values, to obtain a stable FHDE model, one may consider an alternative interacting form between dark matter and dark energy or explore alternative IR cutoffs instead of the Hubble horizon, such as the particle horizon, the future event horizon, the Granda--Oliveros horizon, or the \mbox{Ricci horizon.}

\section{Conclusions} \label{sec:5}

Using the fractional entropy originating from fractional quantum mechanics and adopting the Hubble horizon as IR cutoff, a new HDE model has been proposed. It can realize the late time accelerated expansion of the universe for some special values of the model parameter $\alpha$. In this paper, by considering an interaction between the pressureless matter and FHDE, we have analyzed three different IFHDE models: IFHDE-A, IFHDE-B, and IFHDE-C. 

To estimate the parameters for the IFHDE models, we combine the latest observational data, including SNIa, OHD, BAO, and CMB, and obtain the mean and best-fit values for these models. Compared with the $\Lambda$CDM model and by analyzing $\Delta \chi^{2}_{min}$ and $\Delta \mathrm{AIC}$, we find that IFHDE-A is excluded by the observational data, while both the IFHDE-B and IFHDE-C models show some preference from the AIC 
 and fit the observational data comparable to the $\Lambda$CDM model.

Then, using phase space analysis, we find that the IFHDE-B and IFHDE-C models have three critical points, which represent the radiation-dominated epoch, the matter-dominated epoch, and the dark energy-dominated epoch, respectively. By analyzing the stability of these critical points and the phase space trajectories of these models, we find that only the IFHDE-C model can describe the full evolution history of the universe, but the dark energy itself cannot mimic the cosmological constant at the late time acceleration epoch. Subsequently, we adopt the statefinder diagnostic pair to analyze the IFHDE-C model and find that the evolution curve of statefinder pairs for the IFHDE-C model deviates from the $\Lambda$CDM model but converges to the $\Lambda$CDM fixed point and the de Sitter expansion fixed point in the future. This implies that, due to the interaction between dark energy and dark matter, the IFHDE-C model can mimic the $\Lambda$CDM model at late times despite having a non-standard equation of state.

After analyzing the evolution of cosmological parameters of the IFHDE-C model with the best-fit values, we find that the late time evolution of the universe is dominated by dark energy, and a transition from deceleration to acceleration occurs, confirming that this model drives the late time acceleration of the universe.

\acknowledgments{This work was supported by the National Natural Science Foundation of China under Grant Nos. 12405081 and 11865018.}

\end{document}